\documentclass[aps,twocolumn,showpacs,preprintnumbers,prl,amsmath,amssymb,amsfonts,superscriptaddress,floatfix]{revtex4-1}

\usepackage[latin1]{inputenc}
\usepackage{graphics}
\usepackage{color}
\usepackage{amssymb}
\usepackage{amsmath}
\usepackage{overpic}
\usepackage{epstopdf}
\usepackage{bm}
\usepackage{graphicx}
\usepackage{subfigure}

\begin{document}

\title{Intrinsic ultralow lattice thermal conductivity of the unfilled skutterudite FeSb$_3$}

\author{Yuhao Fu}
\affiliation{College of Materials Science and Engineering and Key Laboratory of Automobile Materials of MOE, Jilin University, Changchun 130012, China}
\author{David J. Singh}
%\affiliation{College of Materials Science and Engineering and Key Laboratory of Automobile Materials of MOE, Jilin University, Changchun 130012, China}
\affiliation{Department of Physics and Astronomy, University of Missouri, Columbia, MO 65211-7010 USA}
\author{Wu Li}
\email{wu.li.phys2011@gmail.com}
\affiliation{Institute for Advanced Study, Shenzhen University, 3688 Nanhai Avenue, Shenzhen 518060, China}
\author{Lijun Zhang}
\email{lijun$\_$zhang@jlu.edu.cn}
\affiliation{College of Materials Science and Engineering and Key Laboratory of Automobile Materials of MOE, Jilin University, Changchun 130012, China}

\date{\today}

\begin{abstract}
It has been generally accepted that unfilled skutterudites process high lattice thermal conductivity ($\kappa_{l}$) that can be efficiently reduced upon filling.
Here by using first principles Boltzmann-Peierls transport calculations, 
we find pure skutterudite of FeSb$_3$ with no filler in fact has an intrinsic ultralow $\kappa_{l}$ smaller than that of CoSb$_3$ by one order of magnitude.
The value is even smaller than those of most of the fully filled skutterudites. 
This finding means that with FeSb$_3$ as a reference, filling does not necessarily lower $\kappa_{l}$.  
The ultralow $\kappa_{l}$ of FeSb$_3$ is a consequence of much softened optical phonon branches associated with the weakly bonded Sb$_4$ rings.
They overlap more with heat-carrying acoustic phonons and significantly increase the phase space for three-phonon anharmonic scattering processes.
This provides an alternative non-filling related mechanism for lowering the $\kappa_{l}$ of skutterudites.
\end{abstract}

\maketitle

Skutterudites are an important class of high performance thermoelectrics \cite{morelli1995, sales1996, nolas1999, shi2011, PhysRevB.56.R1650, zhang2006, PhysRevB.56.7376, Rogl2010, PhysRevB.77.094421, han2009, dyck2002, caillat1996, yang2006, PhysRevB.84.235205,
nolas2000, chen2001, lamberton2002, zhang2009, rogl2014, yang_tuning_2016} 
as the embodiment of the Slack's ``electron-crystal phonon-glass'' idea \cite{slack1995}.
The existence of two isosahedron voids in their crystal structures allows for filling in a variety of cations (\textit{e.g.,} rare earth, alkali earth or alkali metals).
This offers dual advantages for good thermoelectrics: first, according to the Zintl concept 
the additional electrons transferred from the electropositive fillers to the CoSb$_3$ framework make possible flexible control of $n$-type doping \cite{shi2011, PhysRevB.56.7376, Rogl2010, han2009, dyck2002,
nolas2000, chen2001, lamberton2002, zhao2006, pei2006}, and provides compensating change to the $p$-type doping with Co replaced by electron deficient Fe \cite{morelli1995, sales1996, PhysRevB.56.R1650, Rogl2010, PhysRevB.84.235205, qiu2011}. 
Second, and more importantly, filling strongly lowers lattice thermal conductivity ($\kappa_{l}$) \cite{morelli1995, sales1996, nolas1999, shi2011, PhysRevB.56.7376, Rogl2010, han2009, dyck2002,
lamberton2002, zhao2006, nolas2000, zhang2009, rogl2014} and optimizing the filling to lower $\kappa_{l}$ both in terms of filling fraction and by using suitable mixtures of filled cations plays a central role in the optimization of high performance skutterudite thermoelectrics \cite{shi2011}.

The physical mechanism responsible for the reduction of $\kappa_{l}$ in filled skutterudites remains elusive after two decades of intensive research.
There are several debated aspects about the nature and role of the vibrations associated with the filled ``rattling'' atoms:
(i) whether the motion of the rattling atoms is incoherent and non-correlated \cite{slack1994, slack1995, Keppens1998, PhysRevLett.90.135505} or coherently couples with the host framework \cite{PhysRevB.76.140301, koza2008};
(ii) whether there exists anharmonic interaction between the localized rattling modes and the propagating phonons of the host framework \cite{PhysRevB.61.R9209, yamakage2009, PhysRevB.81.134301, PhysRevB.89.184304, PhysRevB.91.144304}; 
(iii) whether the reduction of $\kappa_{l}$ originates from the energy dissipation caused by the resonant scattering of the rattling atoms \cite{slack1994, slack1995, sales1996, PhysRevB.56.15081} or the enhanced conventional anharmonic (Umklapp) scattering processes \cite{koza2008, lee2006, PhysRevB.89.184304, PhysRevB.91.144304}. 

Despite controversy over the mechanism, there is consensus that the filling should reduce the $\kappa_{l}$ of skutterudites.
In this Letter we report the finding via first principles transport calculations that 
the recently reported skutterudite FeSb$_3$ \cite{PhysRevB.91.085410, xing2015, marc1997, PhysRevB.84.064302,  PhysRevB.91.014303, PhysRevB.92.205204} often presumed to be closely related to CoSb$_3$ in fact has an very low $\kappa_{l}$ without filling, 
even lower than those of most of the fully filled skutterudites ($e.g.,$ with Ba, La and Ce).
This ultralow $\kappa_{l}$ in an unfilled skutterudite is a consequence of the much softened optical phonon branches that take the role of the rattling modes in the filled skutterudites.
The emerged low-lying optical phonons
overlap more with the heat-carrying acoustic phonons and increase three-phonon anharmonic scattering channels, thus significantly reducing phonon lifetimes and $\kappa_{l}$. 
This finding demonstrates an unexpected mechanism for the reduction of $\kappa_{l}$ in skutterudites.
 
We perform first principles calculations of $\kappa_{l}$ for FeSb$_3$ and fully filled skutterudites of La/CeFe$_4$Sb$_{12}$ 
by iteratively solving the linearized Boltzmann-Peierls transport equation of phonons with the SHENGBTE package \cite{li2014} (see Supplementary methods for more details).
The equilibrium crystal structures and interatomic force constants (IFCs) are obtained from DFT calculations 
with the plane-wave projector-augmented-wave method \cite{PhysRevB.50.17953}, as implemented in the VASP code \cite{PhysRevB.54.11169}.   
We employ the local density approximation (LDA) as exchange-correlation functional. 
A ferromagnetic configuration for FeSb$_3$ is used, which is the lowest-energy magnetic configuration at the LDA level.
Structural optimization is done with the kinetic energy cutoffs of 350 eV or more and the 8$\times$8$\times$8 $k$-point mesh,
which ensures the residual forces smaller than 1x10$^{-4}$ eV/\AA.
The resulted equilibrium lattice constants are slightly smaller than the experimental data (by 2.03\%, 1.78\%, and 2.00\% for FeSb$_3$, LaFe$_4$Sb$_{12}$ and CeFe$_4$Sb$_{12}$, respectively) as in Supplementary Table S1.
The agreements are reasonably good by consideration of the usual underestimation of lattice constants in the DFT-LDA calculations. 
The harmonic and third-order anharmonic IFCs are calculated by using the real-space supercell approach \cite{phonopy,li2014}, in a 3$\times$3$\times$3 supercell with a 2$\times$2$\times$2 $k$-point mesh and a 2$\times$2$\times$2 supercell with a 3$\times$3$\times$3 $k$-point mesh, respectively.
The phonon momenta $q$-mesh of 15$\times$15$\times$15 is used in solving the transport equation to ensure $\kappa_{l}$ converged at the 1$\times$10$^{-6}$ W/mK level. 

\begin{figure}[h]
\includegraphics[width=3.5in]{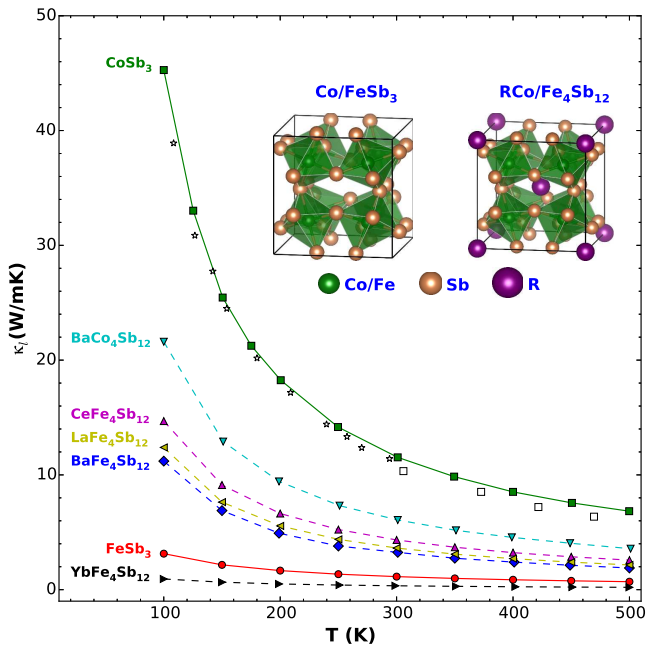}
\centering
\caption{(color online) Calculated temperature dependence of $\kappa_{l}$ (in W/mK) of unfilled skutterudites (FeSb$_{3}$, CoSb$_{3}$ \cite{PhysRevB.90.094302}) and several fully filled skutterudites (LaFe$_{4}$Sb$_{12}$, CeFe$_{4}$Sb$_{12}$, YbFe$_{4}$Sb$_{12}$ \cite{PhysRevB.91.144304}, BaFe$_{4}$Sb$_{12}$ \cite{PhysRevB.91.144304} and BaCo$_{4}$Sb$_{12}$ \cite{PhysRevB.89.184304}). 
The experimental $\kappa_{l}$ for CoSb$_{3}$ (open symbols) are taken from Morelli \textit{et al.} (stars) \cite{PhysRevB.51.9622} and Caillat \textit{et al.} (squares) \cite{caillat1996}. 
The inset shows crystal structures of unfilled and fully filled skutterudites.}
\label{thermalConductivity}
\end{figure}

Fig.\ \ref{thermalConductivity} shows calculated $\kappa_{l}$ as the function of temperature for unfilled FeSb$_{3}$ and CoSb$_{3}$ \cite{PhysRevB.90.094302}, 
as well as fully filled skutterudites of LaFe$_{4}$Sb$_{12}$, CeFe$_{4}$Sb$_{12}$, YbFe$_{4}$Sb$_{12}$ \cite{PhysRevB.91.144304}, BaFe$_{4}$Sb$_{12}$ \cite{PhysRevB.91.144304} and BaCo$_{3}$Sb$_{12}$ \cite{PhysRevB.89.184304}.
The excellent agreement between theoretical results of CoSb$_{3}$ and available experimental data \cite{PhysRevB.51.9622, caillat1996} strongly indicate the validity of our calculations.
Surprisingly, we find that FeSb$_{3}$ exhibits a quite low $\kappa_{l}$ of 1.14 W/mK at 300 K, about one order of magnitude lower than 11.6 W/mK of CoSb$_{3}$.
In the whole temperature range the $\kappa_{l}$ of FeSb$_{3}$ is apparently much lower than the values of the most filled skutterudites (by more than two-third).
The only exception is YbFe$_{4}$Sb$_{12}$ that owes the lowest theoretical $\kappa_{l}$ among reported filled skutterudites \cite{PhysRevB.91.144304}.       
%The $\kappa_{l}$ of FeSb$_{3}$ show a weak temperature dependence, resembling that of YbFe$_{4}$Sb$_{12}$.
 
\begin{figure}[h]
\includegraphics[width=3.5in]{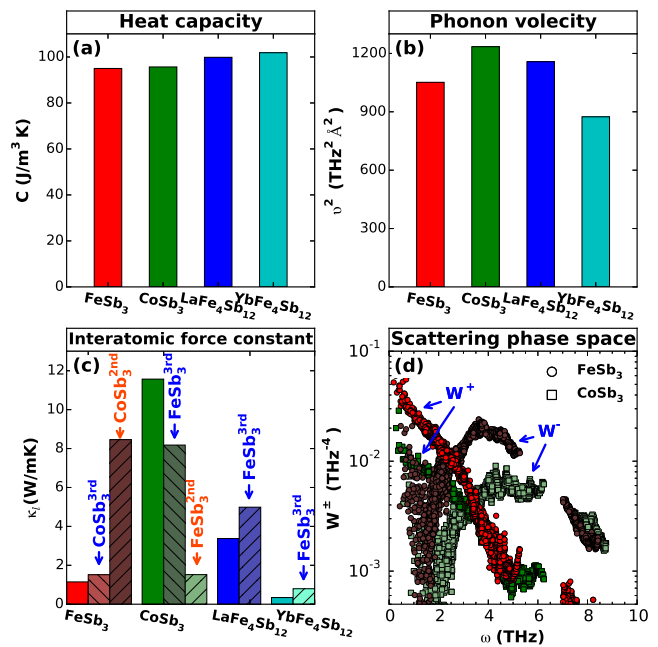}
\centering
\caption{(color online) Analysis of key factors that may affect $\kappa_{l}$ for FeSb$_{3}$, CoSb$_{3}$ and the filled skutterudites of La/YbFe$_{4}$Sb$_{12}$: (a) heat capacity, (b) phonon velocity, (c) interatomic force constant and (d) scattering phase space. 
In (b) the phonon velocity is calculated by averaging the group velocities of the long-wavelength (both transverse and longitudinal) acoustic phonons along different phonon momentum directions.
In (c) the notation of $M^{nth}$ represents that the $nth$-order IFCs of the compound M is used to replace the original IFCs when calculating the room-temperature $\kappa_{l}$.}
\label{analysis}
\end{figure}

We next elucidate the reason why FeSb$_{3}$ has such an ultralow $\kappa_{l}$.  
The physical factors that may affect $\kappa_{l}$ (see Supplementary Eq. S1) include heat capacity $C_\lambda$, phonon velocity $\upsilon_{\lambda}$ and phonon lifetime $\tau_{\lambda}$ of each phonon mode $\lambda$. 
Figs. \ref{analysis}a and \ref{analysis}b show the calculated room-temperature heat capacity 
and the averaged group velocity over the long-wavelength acoustic phonons contributing predominately to heat-carrying, respectively.
The difference in the heat capacities of FeSb$_3$ and CoSb$_3$ is negligibly small (less than 0.7\%) 
and their values are $\sim$7\% lower than those of La/YbFe$_4$Sb$_{12}$.
This is almost exactly as expected from the Dulong-Petit law, in accord with experiments (see Supplementary Fig. S1).
The square of averaged phonon velocity for FeSb$_3$ is about 17\% and 10\% lower than that of CoSb$_3$ and LaFe$_{4}$Sb$_{12}$, respectively.
This originates from the substantially reduced frequencies and velocities of the acoustic phonons in FeSb$_3$ (see below).  
However, the influences of these two factors are far from enough to explain the above large discrepancies in $\kappa_{l}$. 
Therefore it must be the phonon lifetime that plays a central role in reducing $\kappa_{l}$ of FeSb$_{3}$.  

\begin{figure}[h]
\includegraphics[width=3.5in]{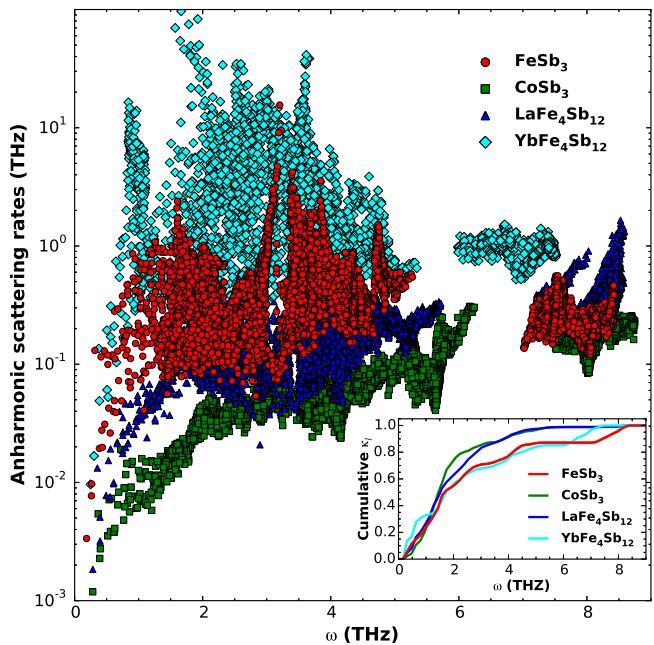}
\centering
\caption{(color online) Calculated anharmonic scattering rates of FeSb$_{3}$ (red circles), CoSb$_{3}$ (green squares), LaFe$_{4}$Sb$_{12}$ (blue triangles) and YbFe$_{4}$Sb$_{12}$ (cyan diamonds) at 300 K. The normalized cumulative $\kappa_{l}$ as the function of $\omega$ is shown in the inset.}
\label{anharmonicSRs}
\end{figure}

The intrinsic phonon lifetimes of crystalline materials are primarily dominated by the three-phonon anharmonic scattering processes. 
%(the contribution from the isotopic disorder scattering is found to be more than two-order smaller).
Fig.\ \ref{anharmonicSRs} shows the calculated anharmonic scattering rates at 300 K.  
As seen, the scattering rates of FeSb$_{3}$ are nearly one order of magnitude larger than those of CoSb$_{3}$ in the low and intermediate frequency ($\omega$) regions (below 5 THz). 
Clearly at most of $\omega$ the rates of FeSb$_{3}$ are much higher than those of LaFe$_4$Sb$_{12}$, though lower than those of YbFe$_4$Sb$_{12}$.
Therefore the observed differences in the anharmonic scattering processes that limit phonon lifetimes indeed account for the discrepancies in $\kappa_{l}$.
Note that the anharmonic scattering rates of FeSb$_{3}$ are remarkably enhanced in a wide intermediate frequency range between $\sim$1 and 5 THz.
The behavior is similar to those in La/YbFe$_4$Sb$_{12}$ and other filled skutterudites \cite{PhysRevB.91.144304, PhysRevB.91.144304, PhysRevB.89.184304}.
This affects contributions of phonon modes to $\kappa_{l}$, 
as indicated by the cumulative plot of $\kappa_{l}$ ($\kappa^c_{l}$) that represents the fraction of heat carried by the phonons with less frequencies than $\omega$ (inset of Fig. \ref{anharmonicSRs}).
The higher anharmonic scattering rates correspond to the smaller lifetimes, and thus the less contributions of the phonons to $\kappa_{l}$.
While for CoSb$_{3}$ $\kappa^c_{l}$ increase rapidly with $\omega$ and the phonons below 2 THz have already contributed to $\sim$80\% of $\kappa_{l}$, 
$\kappa^c_{l}$ of FeSb$_{3}$ show a much slower increase and the phonons below 2 THz only contribute to $\sim$50\% of $\kappa_{l}$.
%This is attributed to the high anharmonic scattering rates of FeSb$_{3}$ in the intermediate frequency range.
The behavior of $\kappa^c_{l}$ of FeSb$_{3}$ resembles those of filled skutterudites, especially YbFe$_4$Sb$_{12}$.
%implying the similar mechanism underlying their low $\kappa_{l}$.  

We further investigate the roles of harmonic and third-order anharmonic IFCs in enhancing three-phonon anharmonic scattering processes and reducing $\kappa_{l}$ in FeSb$_{3}$. 
In particular we perform calculations of $\kappa_{l}$ by deliberately interchanging the harmonic/anharmonic IFCs between two different compounds, 
as shown in Fig.\ \ref{analysis}c.
For FeSb$_{3}$, when replacing the anharmonic IFCs by the ones from CoSb$_{3}$ and remaining the other quantities unchanged, 
we find that $\kappa_{l}$ increases by $\sim$30\%, 
whereas $\kappa_{l}$ of CoSb$_{3}$ decreases by $\sim$30\% when using the anharmonic IFCs from FeSb$_{3}$.
The anharmonic scattering rates are generally proportional to the square of the anharmonic IFCs (Eq. S3 and S4). 
The result means the anharmonic IFCs of FeSb$_{3}$ are larger than those of CoSb$_{3}$, corresponding to the higher scattering rates in FeSb$_{3}$, 
but far from enough to account for its one-order lower $\kappa_{l}$.
For the filled skutterudites of LaFe$_4$Sb$_{12}$ and YbFe$_4$Sb$_{12}$, when using the anharmonic IFCs from FeSb$_{3}$, 
the resulted $\kappa_{l}$ show $\sim$50\% and $\sim$130\% increases, respectively.
This indicates the anharmonic IFCs of FeSb$_{3}$ are smaller, corresponding to the lower scattering rates, 
which conflicts with the smaller $\kappa_{l}$ of FeSb$_{3}$ than that of LaFe$_4$Sb$_{12}$.
When we interchange the harmonic IFCs between FeSb$_3$ and CoSb$_3$,
we find $\kappa_{l}$ of FeSb$_{3}$ increases by about 8 times, and $\kappa_{l}$ of CoSb$_{3}$ decreases by almost the same amount.
The changes accord well with the discrepancy of $\kappa_{l}$ between FeSb$_3$ and CoSb$_3$.
From these results, we can conclude that the main factor responsible for the enhanced anharmonic scattering in FeSb$_{3}$ is the harmonic IFCs, rather than the third-order anharmonic IFCs. 

\begin{figure}[h]
\includegraphics[width=3.5in]{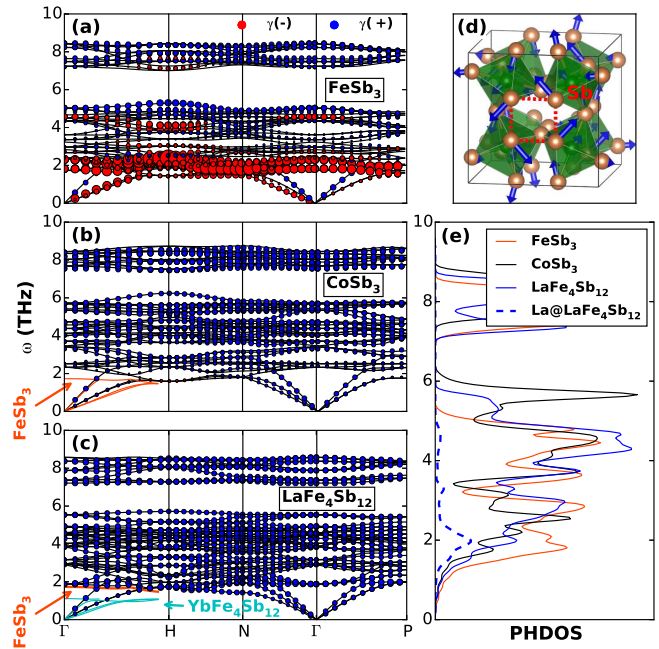}
\centering
\caption{(color online) (a,b,c) Calculated phonon dispersion curves of FeSb$_{3}$, CoSb$_{3}$ and LaFe$_4$Sb$_{12}$. 
The size of circle represents the magnitude of Gr$\ddot u$neisen parameter ($\gamma$) for each mode, 
and the positive $\gamma$ is shown in blue, negative in red. 
In (b) and (c) some phonon branches taken from FeSb$_3$ (in orange) and YbFe$_4$Sb$_{12}$ (in cyan) are shown for comparison.
(d) Eigenvector of the lowest optical mode at the $\Gamma$ point for FeSb$_3$. 
The Sb$_4$ ring is indicated with red dash lines. 
(e) shows the (projected) phonon density of states (PHDOS).}
\label{phononDispersion}
\end{figure}

The specific way of the harmonic IFCs affecting the anharmonic scattering processes is through phonon frequencies and eigenvectors (Eq. S3 and S4).
The phonon eigenvectors contribute to the three-phonon scattering matrix elements. 
Usually for the same class of materials, the changes of eigenvectors with the varied harmonic IFCs are not as large as the changes of frequencies.
It is thus reasonably to assume the scattering matrix elements do not change substantially here.
The action of phonon frequencies on the anharmonic scattering is embodied by the three-phonon scattering phase space W$^{\pm}$ (see Eq. S5), 
which depicts available three-phonon scattering channels among all modes.  
It consists of two components corresponding to absorption (W$^{+}$, two phonons merging into one) and emission (W$^{-}$, one splitting into two) processes, respectively.
As show in Fig.\ \ref{analysis}d, 
the W$^{+}$ of FeSb$_3$ are distinctly larger than those of CoSb$_3$ by several times in the low frequency ($<$ 3.5 THz) region,
and the W$^{-}$ of FeSb$_3$ show much larger values in the wide frequency region below 5 THz as well.
The origin of the significantly enhanced W$^{\pm}$ of FeSb$_3$ lies in its distinct phonon spectrum.
Comparing with that of CoSb$_3$ (Fig.\ \ref{phononDispersion}b), the phonon spectrum of FeSb$_3$ (Fig.\ \ref{phononDispersion}a) shows a clear softening for the lowest optical branch (down to below 2 THz).
This optical branch is even lower in frequency than the La-derived rattling mode in LaFe$_4$Sb$_{12}$ (Fig.\ \ref{phononDispersion}c), 
and is not far away from the extremely low frequency Yb-derived mode in YbFe$_4$Sb$_{12}$ (cyan curve in Fig.\ \ref{phononDispersion}c).    
%On the one hand, the appearance of this low frequency optical branch in FeSb$_3$ makes the acoustic phonons, especially the transverse modes, lower in frequency and velocity (Fig.\ \ref{phononDispersion}b).
Actually it is not only the lowest optical branch, but several adjacent upper optical branches that become softened in FeSb$_3$. 
This is unambiguously reflected by a sharp PHDOS peak appearing around 2 THz (Fig.\ \ref{phononDispersion}e).
It is located in the region similar to that of the La-derived rattling modes in LaFe$_4$Sb$_{12}$ (blue dash line in Fig.\ \ref{phononDispersion}e).    
The low-lying optical phonons overlaps more with the acoustic branches, appreciably increases the phase space W$^{\pm}$ for three-phonon scattering processes (Fig.\ \ref{analysis}d).
This reduces significantly phonon lifetimes, which is the main root cause for the ultralow $\kappa_{l}$ of FeSb$_3$.

\begin{figure}[h]
\includegraphics[width=3.5in]{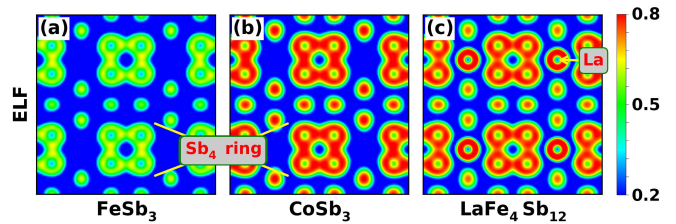}
\centering
\caption{(color online) Contour plot of electron localization function within the (100) plane of (a) FeSb$_{3}$, (b) CoSb$_{3}$ and (c) LaFe$_4$Sb$_{12}$.}
\label{elf}
\end{figure}

Fig.\ \ref{phononDispersion}d shows the vibration pattern of the lowest optical phonon mode of FeSb$_3$.
It involves torsion of the Sb$_4$ ring, a typical quasi-molecular motif in skutterudites.
The softening of this optical mode in FeSb$_3$ originates from the weaker Sb-Sb bonds of the Sb$_4$ ring, 
as demonstrated by the electron localization function contour plot in Fig. \ref{elf}.
Clearly the electrons in the Sb$_4$ ring of FeSb$_3$ are much less localized than the case of CoSb$_3$.
This indicates the rather weaker Sb-Sb bonds in FeSb$_3$, as expected from its electron deficient nature.
This is consistent with the fact that the Young modulus of FeSb$_{3}$ is smaller than that of CoSb$_{3}$ \cite{PhysRevB.91.014303, PhysRevB.84.064302}.
As the result, the phonon modes that are mainly dominated by Sb atoms (below 6 THz) show general softening in FeSb$_{3}$ (Fig.\ \ref{phononDispersion}e). 
The softening also occurs to the acoustic phonons, especially the transverse modes (Fig.\ \ref{phononDispersion}b), 
which leads to the moderately reduced averaged phonon velocity of FeSb$_{3}$ in Fig. \ref{analysis}b.   
In addition to the phonon softening, the weaker Sb-Sb bonds in FeSb$_3$ results in abnormal Gr$\ddot u$neisen parameters ($\gamma$) of phonons (Fig.\ \ref{phononDispersion}a).
For the low-lying optical branches and the transverse acoustic phonons, the values of $\gamma$ are negative (in red) and quite large in magnitude. 
Such phonon modes with the large magnitude $\gamma$ in principle facilitate high lattice anharmonicity \cite{PhysRev.98.1751} and thus low $\kappa_{l}$.
The negative sign of $\gamma$ for the low-lying optical branches implies that they will be further softened under contraction. 
This is expected to cause the more overlapping with acoustic phonons, more enhanced W$^{\pm}$ and thus even lower $\kappa_{l}$ at high pressures.  

For the filled skutterudites, the electrons of filler transfer to the host framework following the Zintl behavior.
These electrons primarily distribute on the Sb$_4$ ring, 
which considerably strengthens the Sb-Sb covalent bonds, as indicated in Fig. \ref{elf}c for the LaFe$_4$Sb$_{12}$ case. 
This is consistent with the band structure of skutterudites, which shows a light Sb derived band at the top of the valence bands \cite{PhysRevB.50.11235}.
This may also explain why filled Fe-based skutterudites are easier to form than FeSb$_3$ usually stabilized in films \cite{marc1997, PhysRevB.84.064302, PhysRevB.91.085410}.
If one considers only the fact that the strengthened Sb-Sb bonds after La filling lift the Sb-derived optical branches in LaFe$_4$Sb$_{12}$,
an increase of $\kappa_{l}$ is expected. 
In fact, the La filler derived rattling modes take the role to remarkably enhance W$^{\pm}$. 
As the result $\kappa_{l}$ of LaFe$_4$Sb$_{12}$ is still much lower than that of CoSb$_3$, and only about three times larger than that of FeSb$_3$.
For the case of YbFe$_4$Sb$_{12}$, the Yb derived even lower frequency and rather flat optical phonon branches increase W$^{\pm}$ more significantly, 
resulting in the more reduced $\kappa_{l}$ \cite{PhysRevB.91.144304}. 

To summarize, we report a discovery of ultralow lattice thermal conductivity ($\kappa_{l}$) of pure skutterudite FeSb$_3$ with no filler by using first principles Boltzmann-Peierls transport simulations.
The calculated $\kappa_{l}$ is only 1.14 W/mK at room temperature, one order of magnitude lower than that of CoSb$_3$.
It is even lower than the values of the most fully filled skutterudites.
This is in contrast to the generally accepted approach where filling is used for reduction of $\kappa_{l}$ in skutterudites.
The origin of the ultralow $\kappa_{l}$ is attributed to the remarkably softened optical phonon branches associated with the weaker Sb-Sb bonds in FeSb$_3$ owing to its electron deficient nature. 
These low frequency optical phonons, having similar frequencies to those of the rattling modes in filled skutterudites, take the role of the rattling modes.
They overlap more with the heat-carrying acoustic phonons and increase significantly the phase space of three-phonon anharmonic scattering processes.
This leads to much reduced phonon lifetimes, and thus very low intrinsic $\kappa_{l}$.
By finding the intrinsic ultralow $\kappa_{l}$ of an unfilled skutterudite, 
our results offer new insight into the still debated mechanism responsible for the reduction of $\kappa_{l}$ upon filling in skutterudites. 
%calling for future experimental verification.
Finally we evaluate the maximum mean free path of phonons for bulk FeSb$_3$ and find the value smaller than 600 nm (Supplementary Fig. S3),
suggesting that the experimentally synthesized FeSb$_3$ films \cite{PhysRevB.84.064302} can be directly applied to verify our theoretical prediction.

The authors acknowledge funding support from the Recruitment Program of Global Youth Experts in China and special funds for talent development in Jilin Province. Part of calculations was performed in the high performance computing center of Jilin University. Work of D.J.S. is supported by S3TEC an Energy Frontier Research Center funded by the Department of Energy, Office of Science, Basic Energy Sciences under award \#DE-SC0001299 / DE-FG02-09ER46577.

\bibliography{ref_skutterudites}

\end{document}

\title{\textbf{Supplemental Material} for ``Intrinsic ultralow lattice thermal conductivity of the unfilled skutterudite FeSb$_3$''}

\author{Yuhao Fu}
\affiliation{College of Materials Science and Engineering and Key Laboratory of Automobile Materials of MOE, Jilin University, Changchun 130012, China}
\author{David J. Singh}
%\affiliation{College of Materials Science and Engineering and Key Laboratory of Automobile Materials of MOE, Jilin University, Changchun 130012, China}
\affiliation{Department of Physics and Astronomy, University of Missouri, Columbia, MO 65211-7010 USA}
\author{Wu Li}
%\email{wu.li.phys2011@gmail.com}
\affiliation{Institute for Advanced Study, Shenzhen University, 3688 Nanhai Avenue, Shenzhen 518060, China}
\author{Lijun Zhang}
%\email{lijun_zhang@jlu.edu.cn}
\affiliation{College of Materials Science and Engineering and Key Laboratory of Automobile Materials of MOE, Jilin University, Changchun 130012, China}

\date{\today}
\maketitle

\section{\textbf{Supplementary Medthods}}

%The skutterudies have the isotopic thermal conductivity. 
The first principles calculations of lattice thermal conductivity ($\kappa_l$) are performed by solving the linearized Boltzmann-Peierls transport equation of phonons \cite{li2014}.
All the contributions from two-phonon and three-phonon scattering processes responsible for intrinsic $\kappa_l$ of crystalline materials are included. 
Briefly the isotopic $\kappa_l$ of cubic skutterudies at temperature T can be represented as the sum of contributions over each phonon mode $\lambda$ (with branch p and wave vector \textbf{q}):

\begin{equation}
\kappa_l\equiv\kappa_l^{\alpha\alpha}=\frac{1}{NV}\sum_{\lambda}C_\lambda\upsilon_{\lambda}^{\alpha}\upsilon_{\lambda}^{\alpha}\tau_{\lambda},
\label{kappa}
\end{equation}

\noindent and

\begin{equation}
C_\lambda=\frac{\partial f_{\lambda}(\omega_{\lambda}, T)}{\partial T},
\label{c}
\end{equation}

\noindent where N is the number of \textbf{q} points uniformly sampled in the phonon Brillouin zone, 
V is the unit cell volume,
$C_\lambda$ is the phonon mode heat capacity, 
$f_{\lambda}(\omega_{\lambda}, T)$ is the Bose-Einstein distribution form that is the function of the phonon frequency $\omega_{\lambda}$ and T, 
$\upsilon_{\lambda}^{\alpha}$ is the phonon group velocity along the $\alpha$ direction,
and $\tau_{\lambda}$ is the phonon lifetime.
In bulk materials without impurities, $\tau_{\lambda}$ is determined by the processes of two-phonon scattering from isotopic disorder and three-phonon anharmonic scattering. 
In the relaxation time approximation, $\tau_{\lambda}$  is equal to a sum of the isotope scattering rate ($\frac{1}{\tau^{iso}}$) and the anharmonic scattering rate ($\frac{1}{\tau^{anh}}$). 
For most of materials, $\frac{1}{\tau^{iso}}$ is at least two-order smaller in magnitude than $\frac{1}{\tau^{anh}}$.
$\frac{1}{\tau^{anh}}$ can be calculated as the sum over $\lambda$ of the three-phonon transition probabilities $\Gamma_{\lambda\lambda^{'}\lambda^{''}}^{\pm}$, 
which can be expressed as \cite{broido2007, PhysRevB.80.125203, li2014, PhysRevB.86.174307}:

\begin{equation}
\Gamma_{\lambda\lambda^{'}\lambda^{''}}^{\pm}=\frac{\hbar\pi}{8N}\{_{f_{{\lambda}^{'}}+f_{{\lambda}^{''}}+1}^{2(f_{{\lambda}^{'}}-f_{{\lambda}^{''}})}\}\frac{\delta(\omega_{\lambda}\pm\omega_{{\lambda}^{'}}-\omega_{{\lambda}^{''}})}{\omega_{\lambda}\omega_{{\lambda}^{'}}\omega_{{\lambda}^{''}}}|V_{\lambda\lambda^{'}\lambda^{''}}^{\pm}|^2,
\label{gamma}
\end{equation}

\noindent and

\begin{equation}
V_{\lambda\lambda^{'}\lambda^{''}}^{\pm}=\sum_{i\in u.c.}\sum_{j,k}\sum_{\alpha\beta\gamma}\frac{e_{\lambda}^{\alpha}(i)e_{p^{'},\pm q^{'}}^{\beta}(j)e_{p^{''},-q^{''}}^{\gamma}(k)}{\sqrt{M_{i}M_{j}M_{k}}}\Phi_{ijk}^{\alpha\beta\gamma},
\label{v}
\end{equation}

\noindent where the upper (lower) row in curly brackets corresponds to the +(-) sign of $\Gamma_{\lambda\lambda^{'}\lambda^{''}}^{\pm}$, which represent three-phonon absorption (two phonons merging into one phonon) and emission (one phonon splitting into two) processes. 
The scattering matrix elements $V_{\lambda\lambda^{'}\lambda^{''}}^{\pm}$ can be evaluated with the normalized eigenvectors $e_{p,q}$ of the three phonons involved 
and the anharmonic interatomic force constants (IFCs) $\Phi_{ijk}^{\alpha\beta\gamma}$ \cite{li2014, PhysRevB.86.174307}. 
The contribution of harmonic phonon frequencies to the anharmonic scattering rates can be represented by the three-phonon scattering phase space W$^{\pm}$ (+ and - signs corresponding to absorption and emission processes).
It is defined as the sum of frequency-containing factors in the expression of three-phonon transition probabilities (Eq.\ \ref{gamma}), and is written as \cite{PhysRevB.89.184304}:
\begin{equation}
W_{\lambda}^{\pm}=\frac{1}{2N}\sum_{\lambda^{'}p^{''}}\{_{f_{{\lambda}^{'}}+f_{{\lambda}^{''}}+1}^{2(f_{{\lambda}^{'}}-f_{{\lambda}^{''}})}\}\frac{\delta(\omega_{\lambda}\pm\omega_{{\lambda}^{'}}-\omega_{{\lambda}^{''}})}{\omega_{\lambda}\omega_{{\lambda}^{'}}\omega_{{\lambda}^{''}}}.
\label{w}
\end{equation}

\pagebreak
\clearpage

%\section{\textbf{Supplementary Figures and Tables}}

\vspace*{\fill}
\begin{figure}[h]
\includegraphics[width=5in]{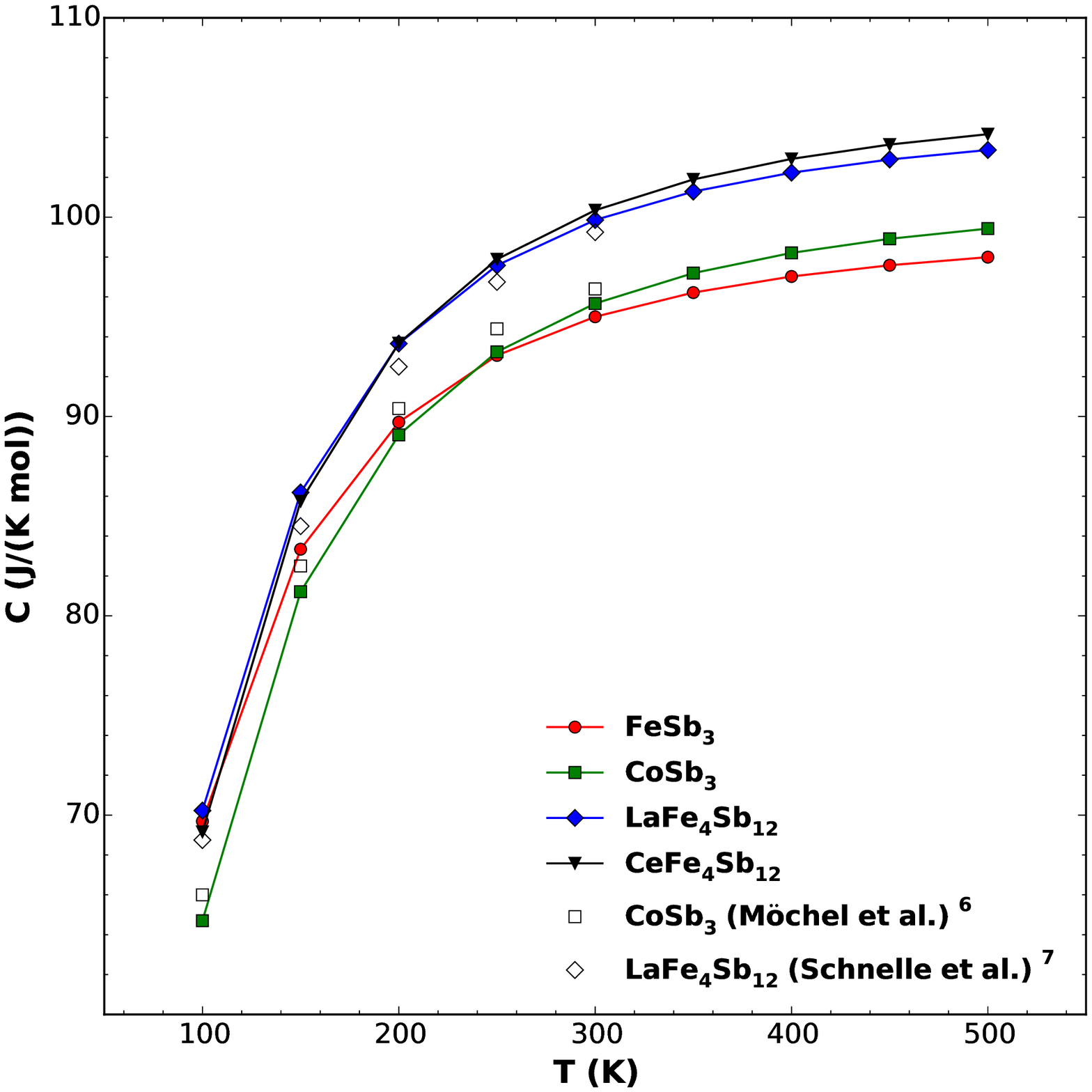}
\centering
\caption{(color online) Calculated heat capacity as the function of temperature for FeSb$_3$, CoSb$_3$, LaFe$_4$Sb$_{12}$ and CeFe$_4$Sb$_{12}$. The available experimental data for CoSb$_3$ \cite{PhysRevB.84.064302} and LaFe$_4$Sb$_{12}$ \cite{PhysRevB.77.094421} are shown for comparison. The agreement between our calculation and the experiments is good. \textbf{In addition to this comparison, the validity of our first principles calculations is also indicated by the comparison of $\kappa_l$ for CoSb$_3$ between our theory and available experiments in Fig. 1 of the main text, as well as the following comparison of phonon spectrum for LaFe$_4$Sb$_{12}$ (Fig. \ref{La}).}}
\label{specificHeat}
\end{figure}
\vfill
\clearpage

\vspace*{\fill}
\begin{figure}[h]
\includegraphics[width=5in]{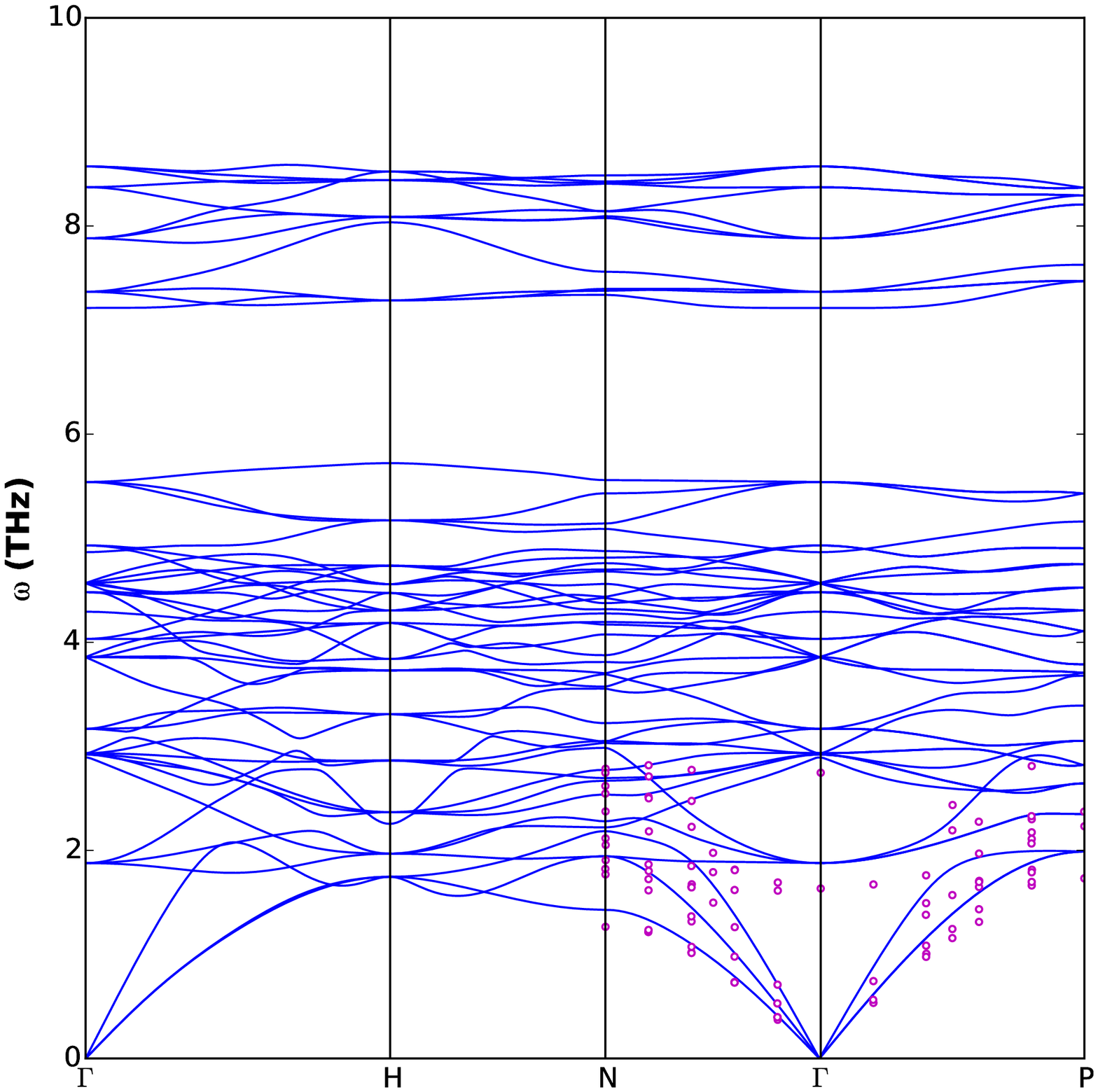}
\centering
\caption{(color online) Calculated phonon dispersion curves of LaFe$_4$Sb$_{12}$. The available experimental data from inelastic neutral scattering measurements \cite{PhysRevB.91.014305} are shown for comparison. The agreement between our calculation and the experiment is reasonably good.}
\label{La}
\end{figure}
\vfill
\clearpage

%\pagebreak
%\clearpage
%\vspace*{\fill}
%\begin{figure}[h]
%\includegraphics[width=5in]{cumulativityThermalOmega.eps}
%\centering
%\caption{(color online) Cumulativity thermal conductivity vs. phonon frequency at 300K.}
%\label{cumulativityThermalOmega}
%\end{figure}
%\vfill
%\clearpage
%
%\pagebreak
%\clearpage
%\vspace*{\fill}
%\begin{figure}[h]
%\includegraphics[width=5in]{pdos.eps}
%\centering
%\caption{(color online) Projected phonon density of states for YbFe$_4$Sb$_{12}$.}
%\label{pdos}
%\end{figure}
%\vfill
%\clearpage
%
%\pagebreak
%\clearpage
%\vspace*{\fill}
%\begin{figure}[h]
%\includegraphics[width=5in]{ifc.eps}
%\centering
%\caption{(color online) Transferability of third-order IFCs among LaFe$_4$Sb$_{12}$, CeFe$_4$Sb$_{12}$ and YbFe$_4$Sb$_{12}$.}
%\label{ifc}
%\end{figure}
%\vfill
%\clearpage
%
%\pagebreak
%\clearpage
%\vspace*{\fill}
%\begin{figure}[h]
%\includegraphics[width=5in]{weigthedPhaseSpace.eps}
%\centering
%\caption{(color online) Calculation weighted phase space and culuative PDOS.}
%\label{weigthedPhaseSpace}
%\end{figure}
%\vfill
%\clearpage
%
%\pagebreak
%\clearpage
%\vspace*{\fill}
%\begin{figure}[h]
%\includegraphics[width=5in]{gruneisen.eps}
%\centering
%\caption{(color online) Calculation gr$\ddot u$neisen parameters of each modes.}
%\label{gruneisen}
%\end{figure}
%\vfill
%\clearpage
%
\pagebreak
\clearpage
\vspace*{\fill}
\begin{figure}[h]
\includegraphics[width=5in]{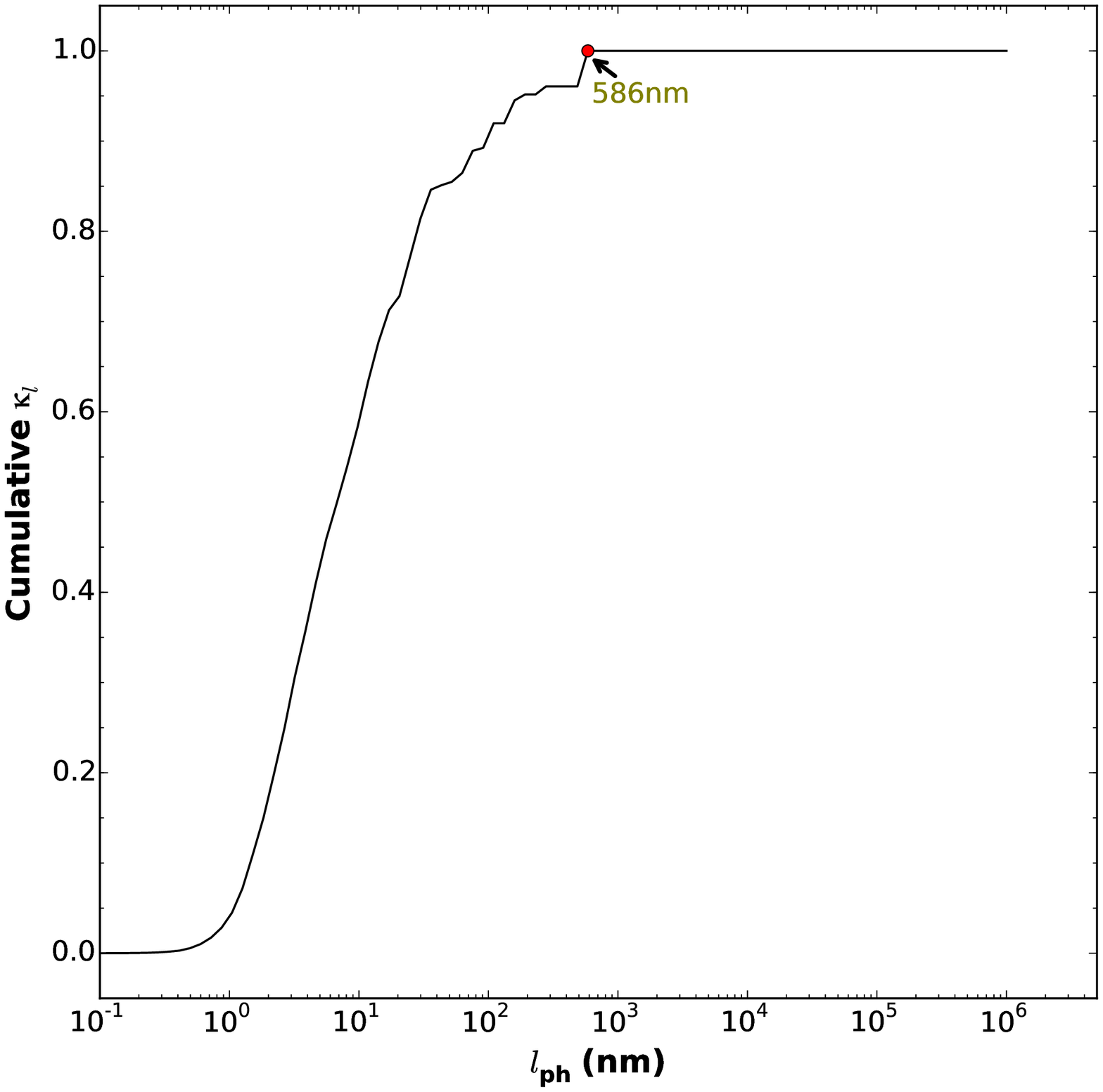}
\centering
\caption{(color online) The normalized cumulative $\kappa_l$ of FeSb$_3$ as a function of the phonon mean free path ($l_{ph}$) at 300 K. The maximum $l_{ph}$ among all the modes, which corresponds to the onset point of the cumulative $\kappa_l$ becoming constant (equal to 1.0), is indicated. 
\textbf{The purpose of this calculation is to provide a reference to future experimental measurements for how thick thin-film samples are required to take on the intrinsic $\kappa_l$ of bulk FeSb$_3$ we calculated here. According to our result, the maximum $l_{ph}$ is less than 600 nm. The sample size above this value is needed to safely avoid surface and grain boundary scatterings. Therefore the experimentally synthesized FeSb$_3$ films with the thicknesses of 1.0-1.5 $\mu m$ \cite{PhysRevB.84.064302} may be directly applied to measure the intrinsic $\kappa_l$ and verify our theoretical prediction.}}
\label{cumulativityThermal}
\end{figure}
\vfill
\clearpage

\pagebreak
\clearpage
\vspace*{\fill}
\begin{table}[h]
  \caption{Optimized structure parameters by total energy minimization for FeSb$_3$, CoSb$_3$, LaFe$_4$Sb$_{12}$ and CeFe$_4$Sb$_{12}$, compared with available experimental data \cite{PhysRevB.91.085410, caillat1996, qiu2011}.
\vspace{0.5cm}  
  }
  
% title of Table
\centering
%\resizebox{17cm}{!}{
%  \begin{minipage}[!b]{20cm}
% used for centering table
\begin{tabular}{lccccc}
% centered columns (6 columns)
\hline\hline
%inserts double horizontal lines
 & & Lattice parameter & \multicolumn{2}{c}{Fractional coordinates}    \\
 & & a & x & y \\
 \hline
\textbf{FeSb$_{3}$} & Theory & 8.968 & 0.329 & 0.161 \\                                                                                                                      
                                    & Exp$^{9}$ & 9.154 & 0.334 & 0.158 \\
 \hline
\textbf{CoSb$_{3}$} & Theory & 8.918 & 0.332 & 0.160 \\
                                    & Exp$^{10}$ & 9.0345 \\
 \hline
\textbf{LaFe$_{4}$Sb$_{12}$} & Theory & 8.985 & 0.335 & 0.164 \\
                                                      & Exp$^{11}$ & 9.148 \\
 \hline
\textbf{CeFe$_{4}$Sb$_{12}$} & Theory & 8.957& 0.333 & 0.164 \\
                                                       & Exp$^{11}$ & 9.140 \\
\hline
% [1ex] adds vertical space
%inserts single line
\end{tabular}
%  \end{minipage}
%  }
\label{latticeParameter}
% is used to refer this table in the text
\end{table}
\vfill
\clearpage

\bibliography{ref_skutterudites}